\documentclass[12pt,epsf]{article}
\usepackage{amsmath}
\usepackage{amssymb}
\begin{document}
\def\a{\alpha}
\def\b{\beta}
\def\c{\varepsilon}
\def\d{\delta}
\def\e{\epsilon}
\def\f{\phi}
\def\g{\gamma}
\def\h{\theta}
\def\k{\kappa}
\def\l{\lambda}
\def\m{\mu}
\def\n{\nu}
\def\p{\psi}
\def\q{\partial}
\def\r{\rho}
\def\s{\sigma}
\def\t{\tau}
\def\u{\upsilon}
\def\v{\varphi}
\def\w{\omega}
\def\x{\xi}
\def\y{\eta}
\def\z{\zeta}
\def\D{{\mit \Delta}}
\def\G{\Gamma}
\def\H{\Theta}
\def\L{\Lambda}
\def\F{\Phi}
\def\P{\Psi}

\def\S{\Sigma}

\def\o{\over}
\def\beq{\begin{eqnarray}}
\def\eeq{\end{eqnarray}}
\newcommand{\gsim}{ \mathop{}_{\textstyle \sim}^{\textstyle >} }
\newcommand{\lsim}{ \mathop{}_{\textstyle \sim}^{\textstyle <} }
\newcommand{\vev}[1]{ \left\langle {#1} \right\rangle }
\newcommand{\bra}[1]{ \langle {#1} | }
\newcommand{\ket}[1]{ | {#1} \rangle }
\newcommand{\EV}{ {\rm eV} }
\newcommand{\KEV}{ {\rm keV} }
\newcommand{\MEV}{ {\rm MeV} }
\newcommand{\GEV}{ {\rm GeV} }
\newcommand{\TEV}{ {\rm TeV} }
\def\slash#1{\ooalign{\hfil/\hfil\crcr$#1$}}
\def\diag{\mathop{\rm diag}\nolimits}
\def\Spin{\mathop{\rm Spin}}
\def\SO{\mathop{\rm SO}}
\def\O{\mathop{\rm O}}
\def\SU{\mathop{\rm SU}}
\def\U{\mathop{\rm U}}
\def\Sp{\mathop{\rm Sp}}
\def\SL{\mathop{\rm SL}}
\def\tr{\mathop{\rm tr}}

\baselineskip 0.7cm

\begin{titlepage}

\begin{flushright}
UCB-PTH-08/10
\end{flushright}

\vskip 1.35cm
\begin{center}
{\Large \bf Dangerous Liouville Wave}

{\large \bf --- exactly marginal but non-conformal deformation}
\vskip 2.2cm

{\large Chiu Man Ho\footnote{E-mail: cmho@berkeley.edu} and Yu Nakayama\footnote{E-mail: nakayama@berkeley.edu}}

\vspace{1cm}

\textit{ Department of Physics, University of California, Berkeley,
CA 94720 \\ Theoretical Physics Group, Lawrence Berkeley National
Laboratory, Berkeley, CA 94720}

\vskip 2.5cm

\abstract{We give a non-trivially interacting field theory example
of scale invariant but non-conformal field theory. The model is
based on the exactly solvable Liouville field theory coupled with
free scalars deformed by an exactly marginal operator. We show
non-vanishing of the trace of the energy-momentum tensor by using
the quantum Schwinger-Dyson equation for the Liouville field theory,
which is a sophistication of the quantum higher equations of motion
for the Liouville field theory introduced by Alyosha Zamolodchikov. Possibly
dangerous implications for the super-critical string theory will be
discussed.}

\end{center}
\end{titlepage}

\setcounter{page}{2}


\section{Introduction}
The distinction between the scale invariance and conformal
invariance was certainly an issue at the advent of the latter.
According to a legend \cite{Migdal}, when a provocative question
about the difference between the two was addressed by a Western
physicist at an international conference on scale invariance in
Dubna, a great mathematician, who was a chairman at the session,
literally said ``There is no mathematical difference, but when some
young people want to use a fancy word they call it Conformal
Symmetry". A young brilliant physicist in the Soviet Union suddenly
stood up and yelled ``15 parameters!" but it echoed apparently
unnoticed.

This issue is not so trivial, and the great mathematician was in
some sense correct from the viewpoint of empirical science because
we do not know any good physical examples of scale-invariant but
non-conformal field theories in four-dimension. In two-dimension,
his claim is even mathematically true because, as later discovered
\cite{Mack1}\cite{Zamolodchikov:1986gt}\cite{Polchinski:1987dy}, one
can give a proof of the equivalence between the two notions under
certain conditions such as unitarity.

Today, the question whether the conformal symmetry is a fancy
alternative word for the scale invariance is a hot topic in high
energy phenomenology. Followed by a seminal work by H.~Georgi
\cite{Georgi:2007ek}, many works have been done to study a possible
existence of a scale invariant (but non-conformal) hidden sector in
our real world and experimental evidence for such ``unparticle
physics", which is spectacular in many cases. A very few authors
have recognized the difference between the scale invariance and the
conformal invariance in this context, and we have stressed the severe unitarity bound
constraint coming from the latter in
\cite{Nakayama:2007qu}\cite{Grinstein:2008qk}. Given the theoretical
situation above, the experimental discovery of scale invariant but
non-conformal ``unparticle" would be a supreme surprise in
theoretical physics.

As we mentioned, quantum examples of scale invariant but
non-conformal field theory are very scarce (see e.g.
\cite{Hull:1985rc} for a notable exception). \footnote{In
\cite{Treiman:1986ep}, other classical examples of scale invariant
field theory in four-dimension without conformal invariance are
given. However, the scale invariance is spontaneously broken there.}
In this paper, we add a new two-dimensional example of such based on
the Liouville field theory. The model is fully quantized by virtue
of the exact solvability of the Liouville field theory. Although our
model is not unitary as can be inferred from the general ``proof" of
the equivalence between scale invariance and conformal invariance in
two-dimension, it may be applied to the world-sheet formulation of
the perturbative string theory.

The organization of the paper is as follows. In section 2, we review
the relation between scale invariance and conformal invariance from
the viewpoint of conserved currents. In section 3, we introduce a
class of classical examples of scale invariant but non-conformal
field theories in two-dimension. In section 4, we investigate a
quantum version of such a model based on the Liouville field theory.
The quantum Schwinger-Dyson equation in the Liouville field theory,
which is crucial to understand the violation of the conformal
symmetry, is thoroughly studied. In section 5, we give some further
discussions of our results.

\section{Scale invariance vs Conformal invariance}
Einstein's special relativity suggests that a basic space-time
symmetry of the quantum field theory (in $d$-dimension) is generated
by the Poincare algebra:
\begin{align}
i[J^{\mu\nu},J^{\rho\sigma}] &=
\eta^{\nu\rho}J^{\mu\sigma}-\eta^{\mu\rho}J^{\nu\sigma} -
\eta^{\sigma\mu}J^{\rho\nu} + \eta^{\sigma\nu}J^{\rho\mu} \cr
i[P^\mu,J^{\rho\sigma}] &= \eta^{\mu\rho}P^{\sigma} -
\eta^{\mu\sigma} P^\rho \cr[P^\mu,P^\nu] &= 0 \ .
\end{align}
For massless scale invariant theory, one can augment this Poincare
algebra by adding the dilatation operator $D$ as
\begin{align}
[P^\mu,D] &= i P^\mu \cr [J^{\mu\nu},D] &= 0  \ .
\end{align}
The generalization of Coleman-Mandula theorem \cite{Haag:1974qh}
asserts (for $d\ge 3$) that the maximally enhanced bosonic symmetry
of the space-time algebra for massless particles is given by the
conformal algebra (plus some internal symmetries):
\begin{align}
[K^\mu,D] &= -iK^\mu \cr [P^\mu,K^\nu] &= 2i\eta^{\mu\nu}D+
2iJ^{\mu\nu} \cr [K^\mu,K^\nu] &= 0 \cr [J^{\rho\sigma},K^\mu] &=
i\eta^{\mu\rho} K^\sigma - i\eta^{\mu\sigma} K^\rho \ ,
\end{align}
where $K^\mu$ generate special conformal transformation.

As is clear from the group theory structure above, the conformal
symmetry demands the scale invariance but the reverse is not
necessarily true: scale invariance does not always imply conformal
invariance.\footnote{See \cite{Wess}\cite{Mack} for earlier references on the interplay between scale invariance and conformal invariance.} A simple example of such theories with scale invariance
but without conformal invariance is a free massless vector field with
no gauge invariance \cite{Coleman:1970je}\cite{Iorio:1996ad}. The
two-dimensional massless vector field in this context was thoroughly
investigated in \cite{Riva:2005gd}.

 However, in reality, every known unitary quantum scale invariant field theory in higher dimension than two is also conformal. The above-mentioned example of free massless vector field with no gauge invariance is not a unitary theory. Furthermore, one can even give a proof of the equivalence between the scale invariance and the conformal invariance for unitary theories with a discrete spectrum in two-dimension \cite{Polchinski:1987dy}.

The distinction between the scale invariance and the conformal
invariance in field theories can be summarized by the properties of
the symmetric energy-momentum tensor $T_{\mu\nu}$. The dilatation
current $S^\mu$ can be generated by
\begin{eqnarray}
S^\mu = x_\nu T^{\nu \mu} + J^\mu  \ ,
\end{eqnarray}
where $J^\mu$ is a so-called virial current. Conservation of the
dilatation current immediately implies
\begin{eqnarray}
T^{\mu}_{\ \mu} = -\partial_{\mu} J^\mu \ .
\end{eqnarray}
Therefore, the necessary and sufficient condition of the scale
invariance is that the energy-momentum tensor is a total divergence.

Moreover, if the virial current itself is a total derivative:
\begin{align}
T^{\mu}_{\ \mu} &= \partial_{\mu}\partial_{\nu} L^{\mu\nu} \ \ \
(d\ge 3) \cr
              &= \partial^\mu\partial_\mu L \ \ \ \  (d =2) \ ,
\end{align}
one can improve the energy-momentum tensor so that it is traceless
(see e.g. \cite{Polchinski:1987dy} for details)
\begin{eqnarray}
\Theta^{\mu}_{\ \mu} = 0 \ .
\end{eqnarray}
By using this improved traceless energy-momentum tensor, one can
construct conserved currents
\begin{eqnarray}
j_v^\mu = v_{\nu} \Theta^{\nu  \mu} \ ,
\end{eqnarray}
where the vector $v^\mu$ satisfies
\begin{eqnarray}
\partial_\mu v_\nu + \partial_\nu v_\mu = \frac{2}{d} \eta_{\mu\nu} \partial_\rho v^{\rho}  \ .
\end{eqnarray}
The currents $j_v^\mu$ generate all the conformal transformation. In
particular, one can obtain the special conformal transformation
associated with $K^\mu$ by taking $v_\mu = \rho_\mu x^\nu x_\nu - 2
x_\mu \rho^\nu x_\nu$, where $\rho_\mu$ is a constant vector parameter.

In this way, the study of the scale invariant but non-conformal
field theory is reduced to the problem whether the virial current is
a total derivative or not. In two-dimension, one can show that
$\langle \Theta^{\mu \ \dagger}_{\ \mu} \Theta^{\mu}_{\ \mu} \rangle
=0$ with scale invariance
\cite{Mack1}\cite{Zamolodchikov:1986gt}\cite{Polchinski:1987dy},
which implies $\Theta^{\mu}_{\ \mu} =0$ (conformal invariance) for a
unitary and compact theory.

\section{Classical Liouville field theory with dangerous perturbation}
Reference \cite{Iorio:1996ad} showed a class of two-dimensional
examples of classical field theories that are scale invariant but
have no conformal invariance. The model is based on the classical
Liouville field theory, so we would like to begin with a brief
review of the conformal invariance of the classical Liouville field
theory. The Liouville field theory\footnote{We use the convention of
\cite{Nakayama:2004vk}.} has the action
\begin{eqnarray}
S_{\rm Liouville} = \frac{1}{4\pi} \int d^2z \left( \partial^\mu\phi
\partial_\mu \phi + 4\pi\mu e^{2b\phi} \right) \ ,
\end{eqnarray}
where the classical limit corresponds to $b\to 0$.

The Liouville equation can be obtained as an equation of motion:
\begin{eqnarray}
\partial_\mu\partial^\mu\phi = 4\pi\mu be^{2b\phi} \ .
\end{eqnarray}
An energy-momentum tensor can be constructed from the Noether
prescription:
\begin{eqnarray}
T_{\mu\nu} = -\partial_\mu\phi \partial_\nu \phi +
\frac{\eta_{\mu\nu}}{2} \left(\partial_\rho\phi \partial^\rho\phi +
4\pi \mu e^{2b\phi} \right) \ .
\end{eqnarray}
Of course, one could improve the energy-momentum tensor at this
point by adding a total derivative $\partial_\mu\partial_\nu \phi -
\eta_{\mu\nu}\partial_\rho\partial^\rho \phi$, but we do not do it
here.

The trace of the energy-momentum tensor can be evaluated by using
the equation of motion as
\begin{align}
T_{\ \mu}^{\mu} &= 4\pi \mu e^{2b\phi} = \frac{1}{b} \partial_\mu
\partial^\mu \phi \ .
\end{align}
Thus, the virial current $J_\mu = -\frac{1}{b} \partial_\mu\phi$ is
a total derivative, and the Liouville field theory is a conformal
field theory. Indeed, one can construct the traceless energy-momentum tensor as
\begin{eqnarray}
\Theta_{\mu\nu} = T_{\mu\nu} +
\frac{1}{b}\left(\partial_\mu\partial_\nu \phi -
\eta_{\mu\nu}\partial^\rho\partial_\rho \phi\right) \ ,
\end{eqnarray}
which will yield a holomorphic energy-momentum tensor\footnote{A
quantum correction will modify the energy-momentum tensor as $T(z) =
-\partial\phi\partial \phi + Q \partial^2 \phi$, where $Q =
b+b^{-1}$. }
\begin{eqnarray}
T(z) \equiv \Theta_{zz}(z) = -\partial\phi\partial \phi +
\frac{1}{b} \partial^2 \phi \ .
\end{eqnarray}

A class of classical scale invariant but non-conformal field
theories is obtained \cite{Iorio:1996ad} by coupling the Liouville
field theory to a sigma model
\begin{eqnarray}
S = \int d^2z \, G^{MN}(X^N) \partial_\mu X_M \partial^\mu X_N +
S_{\rm Liouville} + S_{\rm int}
\end{eqnarray}
by the interaction
\begin{eqnarray}
S_{\rm int} = \frac{\lambda}{4\pi} \int d^2z \, h(X^N)
\partial_\mu\phi \partial^\mu \phi
\end{eqnarray}
with a nontrivial scalar function $h(X^N)$ in the target space. The
model is classically scale invariant with obvious scaling dimensions
$D(X^N) = 0$ and $D(e^{2b\phi}) = 2$.

However, the model has no conformal invariance. To see this, let us
compute the trace of the energy-momentum tensor:
\begin{eqnarray}
T^{\mu}_{\ \mu} = 4\pi\mu  e^{2b\phi} = \frac{1}{b} \partial_\mu
\left[\left(\,1+\lambda h(X^N)\,\right) \partial^\mu\phi\right] \ ,
\end{eqnarray}
which is divergence of the virial current, and, as a consequence,
the theory is expectedly scale invariant. However, the associated
virial current
\begin{eqnarray}
J_\mu =- \frac{1}{b} \left(\,1+\lambda h(X^N)\,\right) \partial_\mu
\phi
\end{eqnarray}
is not a total derivative for non-trivial $h(X^N)$, so the model is
not a conformal field theory.

Before we go on constructing a quantum version of the above scale
invariant but non-conformal field theory, several comments are in order.
\begin{itemize}
    \item The Liouville interaction is crucial. For $\mu=0$, one can recover the conformal invariance by setting $D(\phi) = 0$. Thus, exact treatment of the Liouville interaction would be needed when quantized.
    \item Quantum mechanically, one has to show that $h(X^N)$ has a non-trivial fixed point as well as the target space metric $G^{MN}(X^N)$. One-loop approximation will give you Einstein-dilaton equation coupled with the non-trivial tachyon. The Liouville interaction is very difficult to treat in this approach because it is strongly coupled, and higher $\alpha'$ corrections cannot be neglected. We take a different root to establish the fixed point in the next section.
    \item The model gives a ``counterexample" for the proof of the equivalence between the scale invariance and conformal invariance in two-dimension. Assuming the nontrivial fixed point for $h(X^N)$, we see that the proof fails because of the non-compactness\footnote{The non-compactness of the target space also played a crucial role in the examples of scale invariant but non-conformal field theories studied in \cite{Hull:1985rc}
.} of the target space (especially in the Liouville direction).
\end{itemize}

\section{Quantum Liouville wave}
In this section, we construct a concrete quantum example of scale
invariant but non-conformal field theory based on the model
presented in section 3. We take a sigma model as a flat target-space
with signature $(1,1)$. The action is
\begin{eqnarray}
S = \frac{1}{4\pi}\int d^2z \left(\partial^\mu X^1 \partial_\mu X^1
- \partial^\mu X^0\partial_\mu X^0 \right) + S_{\rm Liouville} +
S_{\rm int} \ .
\end{eqnarray}
The interaction takes a form of the light-cone wave\footnote{The action is not hermitian with our choice of the interaction. However, since our discussion does not depend $v$ as we will see in the following, one can formally perform analytic continuation $v \to iv$ to make the action hermitian. See also footnote \ref{tt} for a related point. In any case, we do not require unitarity, so this is not a primary concern of our construction.}:
\begin{eqnarray}
S_{\rm int} = \frac{\lambda}{4\pi} \int d^2z \left( e^{iv(X^1-X^0)}
\partial^\mu \phi \partial_\mu \phi \right) \ . \label{interaction}
\end{eqnarray}
In other words, we take $h = e^{iv(X^1-X^0)}$.

As before, the trace of the (classical) energy-momentum tensor
\begin{eqnarray}
T^{\mu}_{\ \mu} = \frac{1}{b} \partial^\mu\left[\left(1+\lambda
e^{iv(X^1-X^0)}\right) \partial_\mu\phi\right]
\end{eqnarray}
cannot be improved to be zero. Alternatively, the formerly
holomorphic energy-momentum tensor now becomes

\begin{eqnarray}
T = -\partial X^1 \partial X^1 + \partial X^0 \partial X^0  -
\left(1+\lambda e^{iv(X^1-X^0)}\right)\partial \phi \partial \phi +
Q
\partial^2 \phi \ ,
\end{eqnarray}
which is classically no longer holomorphic
\begin{eqnarray}
\bar{\partial} T = \frac{1}{b}\partial [\,\partial\bar{\partial}
\phi - \pi\mu b e^{2b\phi}\,] \ \neq 0 \ . \label{chl}
\end{eqnarray}

As a consequence, to see the quantum mechanical violation of the
conformal symmetry of this system, one can investigate the following
correlation functions
\begin{align}
& b \left\langle\,  \bar{\partial} T(x_T) \,\,O_1 \cdots
O_N\,\right\rangle \cr =&\left\langle\,
\partial(\partial \bar{\partial} \phi - \pi \mu b e^{2b\phi}
)(x_T)\,\,O_1 \cdots O_N \,\right\rangle  \cr =& \sum_n \frac{1}{n!}
\left\langle
\partial(\partial \bar{\partial} \phi - \pi \mu b
e^{2b\phi})(x_T)\,\,O_1 \cdots O_N\left[\frac{-\lambda}{4\pi} \int
d^2z \,e^{iv(X^1-X^0)}
\partial_\mu\phi \partial^\mu \phi \right]^n \right\rangle_{\lambda
= 0} \ , \label{ser}
\end{align}
where $O_i$ are inserted at $x= x_i$. We have neglected possible
contact terms in the first equality, which do not play any role in
the conformal symmetry breaking.\footnote{Note that the conformal
invariance (or breaking) does not say anything about the structure
of the contact terms. What is relevant in the following, however, is
that for non-zero $\lambda$, we have to integrate the additional
contact terms over the inserted position to obtain non-zero
non-contact terms that will break conformal invariance. } The second
equality is a perturbative series evaluated by the unperturbed
Liouville field path integral. Actually, the perturbative series is
not a formal summation but contains only a single term for each set
of $O_i$ with fixed charge due to the charge conservation for $X^1$
and $X^0$. In the following, we will show that \eqref{ser} does not
vanish so that the conformal invariance is indeed violated quantum
mechanically.

As a side remark, the first equality in \eqref{chl} might seem to
rely on the classical equation of motion and need possible quantum
modifications in the evaluation of \eqref{ser}. However, the
Liouville equation of motion is exact, so for $\lambda=0$, we do not
need any modification.\footnote{The only exception is the case where
$v=0$. In this case, the contact term between the Liouville equation
of motion and the perturbative interaction after integration, which
is actually nothing but a deliberate separation of the Liouville
kinetic term, gives you contribution $\bar{\partial} T =
-\frac{1}{b}\partial(\lambda \partial \bar{\partial} \phi)$. This
can be absorbed by a redefinition of the holomorphic energy-momentum
tensor as $T \to T + \frac{\lambda}{b} \partial^2 \phi$. Note that
this redefinition cannot be done for non-zero $v$ even classically.
} As a perturbative quantum expansion in $\lambda$, order by order
quantum redefinition of the energy-momentum tensor does not recover
the holomorphicity because it is broken at the classical level and
the quantum modification cannot compensate the classical piece, as
long as the classical equation of motion is compatible with the
exact quantization as we will show explicitly.

Even with the Liouville equation of motion $\partial \bar{\partial}
\phi - \pi \mu b e^{2b\phi} =0$ for the unperturbed action $(\lambda
=0)$, the series \eqref{ser} does not generically vanish. The
quantum equation of motion (Schwinger-Dyson equation) possesses a
contact term at $x_i = x_T$:
\begin{eqnarray}
\frac{2}{\pi}\left\langle \, (\partial \bar{\partial} \phi - \pi \mu
b e^{2b\phi})(x_T)\,\,O_1 \cdots O_N\, \right\rangle_{\lambda=0} =
\sum_i \left\langle O_1 \cdots \frac{\delta O_i(x_i)}{\delta
\phi(x_T)} O_N\right\rangle_{\lambda=0} \ .
\end{eqnarray}
Formally, one can obtain the Schwinger-Dyson equation from the
invariance of the path integral measure
\begin{align}
&\int \mathcal{D}\phi \,\,O_1\cdots O_N \,e^{-S} = \int
\mathcal{D}(\phi +\delta\phi) \,\,O_1 \cdots O_N \,e^{-S} \cr \iff&
0 = \int \mathcal{D}\phi \,\,\frac{\delta}{\delta \phi}\left(O_1
\cdots O_N\, e^{-S}\right) \ .
\end{align}
The contact terms in the Schwinger-Dyson equation at $z= x_T$ after
integrating over the inserted position $z$ will give you the failure
of the holomorphicity of the energy-momentum tensor in \eqref{ser}.

In the following, we focus on the contact terms in the Liouville
equation of motion in the Liouville correlation functions denoted by $\langle \cdots \rangle_L$ among the Liouville
primary vertex operators $V_\alpha \sim e^{2\alpha \phi}$. From the
path integral argument, we expect the following identity:
\begin{align}
 \frac{2}{\pi}&\left\langle \,(\partial\bar{\partial} \phi - \pi\mu b e^{2b \phi})(x_T)\,\,e^{2\alpha_1\phi(x_1)}
 \cdots e^{2\alpha_N \phi(x_N)}\, \right\rangle_L
 \cr=& \sum_i 2\alpha_i \delta(x_i-x_T)  \left\langle\, e^{2\alpha_1\phi(x_1)}
 \cdots e^{2\alpha_N \phi(x_N)} \,\right\rangle_L \ . \label{idd}
\end{align}

The quantum treatment of the higher equations of motion in Liouville
field theory was initiated in \cite{Zamolodchikov:2003yb} (see also
\cite{Bertoldi:2004yk}  for a subsequent work). We first introduce
the logarithmic primary operator:
\begin{eqnarray}
V_{0}' = \frac{1}{2} \frac{\partial}{\partial \alpha}
V_{\alpha}|_{\alpha=0} \simeq \phi \ .
\end{eqnarray}
Then, \cite{Zamolodchikov:2003yb} showed that the correlation
function is invariant under the replacement (we recall $L_{-1} =
\partial$)
\begin{eqnarray}
L_{-1}\bar{L}_{-1} V'_{0} = \pi \mu b V_{\alpha = b} \ .
\end{eqnarray}
The derivation of \cite{Zamolodchikov:2003yb} is only valid up to
contact terms. We now show a refinement of his argument to derive
the contact term contributions to the quantum equation of motion.

As in \cite{Zamolodchikov:2003yb}, we concentrate on the three-point
function
\begin{eqnarray}
\left\langle \, L_{-1}\bar{L}_{-1} V'_{\alpha}(x_T)
V_{\alpha_1}(x_1) V_{\alpha_2}(x_2)\,\right \rangle_L
\end{eqnarray}
and study $\alpha \to 0$ limit. The three-point function takes the
form
\begin{align}
 &\left\langle \,V'_\alpha(x_T) V_{\alpha_1} (x_1) V_{\alpha_2} (x_2)\,\right\rangle_L \cr
=& \frac{1}{2} \frac{\partial}{\partial \alpha} \left[
\frac{C(\alpha,\alpha_1,\alpha_2)}{|x_1-x_2|^{2\Delta_1 +
2\Delta_2-2\Delta}|x_T-x_1|^{2\Delta_1+2\Delta -
2\Delta_2}|x_T-x_2|^{2\Delta_2+2\Delta-2\Delta_1}} \right] \ ,
\label{three}
\end{align}
where the conformal weight of the Liouville primary operator
$V_{\alpha}$ is given by $\Delta = \alpha (Q-\alpha)$. The structure
constant $C(\alpha,\alpha_1,\alpha_2)$ of the Liouville field theory
was computed
\cite{Dorn:1994xn}\cite{Zamolodchikov:1995aa}\cite{Teschner:1995yf}
to be
\begin{align}
C(\alpha,\alpha_1,\alpha_2) =& [\pi\mu
\gamma(b^2)b^{2-2b^2}]^{(Q-\alpha-\alpha_1-\alpha_2)/b} \cr \times &
\frac{\Upsilon'(0)\Upsilon(2\alpha)\Upsilon(2\alpha_1)\Upsilon(2\alpha_2)}{\Upsilon(\alpha
+ \alpha_1+\alpha_2-Q)\Upsilon(\alpha + \alpha_1 -
\alpha_2)\Upsilon(\alpha_1 + \alpha_2 - \alpha)\Upsilon(\alpha_2 +
\alpha - \alpha_1)} \ ,
\end{align}
where $\Upsilon(x)$ is defined by
\begin{eqnarray}
\log \Upsilon(x) = \int_0^\infty
\frac{dt}{t}\left[\left(\frac{Q}{2}-x\right)^2e^{-t} -
\frac{\sinh^2(\frac{Q}{2}-x)\frac{t}{2}}{\sinh\frac{bt}{2}\sinh{\frac{t}{2b}}}\right]
\end{eqnarray}
for $ 0<\mathrm{Re}(x) <Q$ and analytically continued to the whole
complex plane. See e.g.
\cite{Zamolodchikov:1995aa}\cite{Nakayama:2004vk} for some
properties of the special functions.

For generic value of $\alpha_1$ and $\alpha_2$, the structure
constant $C(\alpha,\alpha_1,\alpha_2)$ has a simple zero as $\alpha
\to 0$, and only the term with $\partial_\alpha
C(\alpha,\alpha_1,\alpha_2)$ in \eqref{three} contributes as
discussed in \cite{Zamolodchikov:2003yb}. This is consistent with
the contact term contribution that should yield like
$\delta(x_T-x_1) \langle V_{\alpha_1}(x_1) V_{\alpha_2}
(x_2)\rangle_L$, which is non-zero only in the $\alpha_1 \to \alpha_2$
limit (or $\alpha_1 \to Q-\alpha_2$ limit).

 We, thus, take a careful limit of $\alpha \equiv \epsilon \to 0$ and $\alpha_1-\alpha_2 \equiv i\kappa \to 0$.\footnote{The reason for $i$ in $\alpha_1-\alpha_2$ is that we take the physical normalizable Liouville momenta: $\alpha \in \frac{Q}{2} + i\mathbf{R}$. } In this limit, \eqref{three} becomes
\begin{eqnarray}
\frac{\partial}{\partial \epsilon} \left[\frac{2\epsilon}{(\epsilon
+ i\kappa)(\epsilon
-i\kappa)}\left(\frac{S(\alpha_1)}{|x_1-x_2|^{4\Delta_1-2\epsilon
Q}|x_1-x_T|^{2Q\epsilon}|x_2-x_T|^{2Q\epsilon}} +
\mathcal{O}(\epsilon,\kappa) \right)\right] \ , \label{fact}
\end{eqnarray}
where $S(\alpha_1)$ is the two-point function of the Liouville field
theory: $\langle V_{\alpha_1}(1) V_{\alpha_2}(0) \rangle_L =
S(\alpha_1) \pi \delta(i\alpha_1-i\alpha_2)
+\pi\delta(i\alpha_1+i\alpha_2-iQ) $, whose explicit form is given
by
\begin{eqnarray}
S(\alpha) = \frac{(\pi\mu\gamma
\,b^2)^{(Q-2\alpha)/b}}{b^2}\frac{\gamma(2\alpha b
-b^2)}{\gamma(2-2\alpha/b-1/b^2)} \  ,
\end{eqnarray}
where $\gamma(x) = \frac{\Gamma(x)}{\Gamma(1-x)}$.
 We regard the first factor in \eqref{fact} as the delta-function: $\lim_{\epsilon \to 0} \frac{2\epsilon}{(\epsilon + i \kappa)(\epsilon - i\kappa)} = 2\pi \delta(\kappa)$. Then, the derivative with respect to $\epsilon$ gives the logarithmic term
\begin{eqnarray}
2\pi\delta(i\alpha_1-i\alpha_2) S(\alpha_1)|x_1-x_2|^{-4\Delta_1} 2Q \left(\,\log|x_1-x_T|
+ \log|x_2-x_T|\,\right)\ + \cdots , \label{twop}
\end{eqnarray}
where the ellipsis contains only $x_T$ independent terms.

We take the laplacian of \eqref{twop} with $x_T$ from the insertion
of $L_{-1}\bar{L}_{-1} = \partial \bar{\partial}$. By using the
formula $\partial \bar{\partial} \log|z|^2 =\pi \delta(z)$, we
obtain the sought-after contact term:
\begin{align}
& \left\langle\, (L_{-1}\bar{L}_{-1} V'_0 - \pi \mu b V_{b})(x_T)
\,\,V_{\alpha_1}(x_1)V_{\alpha_2}(x_2) \,\right \rangle_L  \cr =&2\pi
\delta(i\alpha_1-i\alpha_2) S(\alpha_1)|x_1-x_2|^{-4\Delta_1} \pi Q \left(
\,\delta(x_1-x_T) + \delta(x_2-x_T)\,\right)  \ . \label{conl}
\end{align}
In this way, we have shown that the contact terms indeed exist in the exact Liouville equation of motion, and from \eqref{ser}, we can now prove that the conformal invariance is broken for nonzero $\lambda$ in the exact quantization of our model. In particular, note that the operator $\partial^\mu\phi \partial_\mu \phi$ inserted in \eqref{ser} can be realized as a specific limit of the Liouville primary operator: $\partial^\mu\phi \partial_\mu \phi = 4:L_{-1} V'_0 \bar{L}_{-1} V'_0:$.

Another suggestive but not complete way to understand the importance
of the contact terms in the conformal symmetry breaking is to
perform partial integration inside the perturbative deformation to
study the insertion of $\int d^2z \phi \partial \bar{\partial}
\phi$. By using the undeformed Liouville equation of motion, it is
equivalent to the insertion of $\int d^2z \phi e^{2b\phi} =\frac12
\int d^2 z \frac{\partial}{\partial\alpha} V_{\alpha}|_{\alpha =
b}$. The above computation directly shows that there exist contact
terms for the vertex insertion $ \frac{\partial}{\partial\alpha}
V_{\alpha}|_{\alpha = b}$, and the integration over the inserted
position $z$ gives a non-contact term contribution to the
energy-momentum tensor insertion.\footnote{We should note, however,
that there is an additional contribution $\int d^2z
\partial^\mu\left[ e^{iv(X^1-X^0)}\right]\phi \partial_\mu \phi$
which cannot be computed in this approach.}

Nevertheless, the limiting procedure is a little bit subtle and one might claim an
objection to the above derivation especially because \eqref{conl} is
different from the Schwinger-Dyson equation from the naive path
integral \eqref{idd}. However, the naive Schwinger-Dyson equation
\eqref{idd} cannot be correct for the exact Liouville correlation
function among $V_{\alpha}$. It is in contradiction with the
reflection symmetry \cite{Zamolodchikov:1995aa} of the Liouville
field theory: $V_{\alpha} \sim S(\alpha) V_{Q-\alpha}$.

To see this, suppose $V_{\alpha} = e^{2\alpha \phi}$ and use the
naive Schwinger-Dyson equation \eqref{idd}:
\begin{eqnarray}
\frac{2}{\pi}\left\langle \, (L_{-1}\bar{L}_{-1} V'_{0} -\pi\mu
bV_{b})(x_T)\,\,V_{\alpha_1} \cdots V_{\alpha_N} \, \right\rangle_L =
2\alpha_1 \delta(x_T-x_1) \langle V_{\alpha_1} \cdots
V_{\alpha_N}\rangle_L + \cdots \label{niv}
\end{eqnarray}
Alternatively, one could replace $V_{\alpha_1}$ with $S(\alpha_1)
V_{Q-\alpha_1}$, and use the Schwinger-Dyson equation, and then
replace $V_{Q-\alpha_1}$ with $S(\alpha_1)^{-1} V_{\alpha_1}$:
\begin{align}
 &\frac{2}{\pi}\left\langle \, (L_{-1}\bar{L}_{-1} V'_{0}-\pi\mu bV_{b})(x_T)\,\,V_{\alpha_1} \cdots V_{\alpha_N}\, \right\rangle_L \cr
 =& \frac{2}{\pi}S(\alpha_1) \left\langle \, (L_{-1}\bar{L}_{-1} V'_{0} -\pi\mu bV_{b})(x_T)\,\,V_{Q- \alpha_1} \cdots V_{\alpha_N}\,\right \rangle_L \cr
=&  2(Q-\alpha_1) \delta(x_T-x_1) \langle V_{\alpha_1} \cdots
V_{\alpha_N}\rangle_L + \cdots \ ,
\end{align}
which is in contradiction with \eqref{niv}.

Any $\alpha$ dependence in the contact term is inconsistent with the
reflection symmetry of the quantum Liouville field theory. The
limiting procedure we showed in the above is the most natural one
consistent with the reflection symmetry. Indeed, the discussion here
suggests a deep insight about the Liouville primary vertex
operator $V_{\alpha}$. It seems quite plausible that the classical
interpretation of $V_{\alpha}$ is not $e^{2\alpha \phi}$, but rather
a mixture $e^{2\alpha \phi} + S(\alpha) e^{2(Q-\alpha)\phi} +
\cdots$.\footnote{This can be also inferred from the analysis of the
mini-superspace reflection amplitudes \cite{Zamolodchikov:1995aa}.}
With the interim substitution of $V_{\alpha} \sim e^{2\alpha \phi} +
S(\alpha) e^{2(Q-\alpha)\phi}$, the path integral approach in
\eqref{idd} agrees with the exact Schwinger-Dyson equation obtained
from the exact three-point function with our limiting procedure.

\subsection{Scale Invariance}

So far, we have discussed that the conformal symmetry is broken due
to the coupling between the Liouville sector and the free boson
sector. Even quantum mechanically, the Schwinger-Dyson equation of
the Liouville field theory demands that the holomorphy of the
energy-momentum tensor is violated. Now the question is whether the
scale invariance is disturbed by this perturbation quantum
mechanically.   We would like to show some arguments that the
interaction \eqref{interaction} is exactly marginal in the sense
that the scale invariance is preserved.

First of all, as a necessary condition, our interaction Lagrangian
has a quantum scaling dimension $D=2$, which gives a first order
perturbative condition for the scale invariance of the theory. To
see higher order corrections, one can focus on the partition
function
\begin{align}
Z_{\lambda} &= \int \mathcal{D} \phi \mathcal{D}X^1\mathcal{D}X^0
\,\,e^{-S} \cr
  &= \sum_{n} \frac{1}{n!}\left\langle\left(\frac{-\lambda}{4\pi}\int d^2z \,\,e^{iv(X^1-X^0)}\partial_\mu\phi \partial^\mu\phi \right)^n \right\rangle_{\lambda=0} \cr
  &=Z_{\lambda=0} \ .
\end{align}
The last equality is due to the charge conservation. From this
formal expression, one might naively conclude that, according to the
general recipe of the conformal perturbation theory, we would not
introduce any regularization or cut-off dependence, and hence the
higher order beta function vanishes because the perturbative
expansion of the partition function itself vanishes. However, in
order to evaluate the beta function, what one has to really study is
the singularity structure of the operator product expansions (OPEs)
inside the formally vanishing perturbative corrections to the
partition function that could be non-zero by adding background
charges at infinity.

To address this question, we take a closer look at the singularity
structure of the of OPEs of the Liouville sector and the sigma model
sector separately. Firstly, in the Liouville sector, it is crucial
to notice that the operator $\partial_\mu\phi \partial^\mu \phi$ is
an exactly marginal deformation to the Liouville field theory: it
just changes the normalization of the kinetic term.\footnote{One
should note that because of the changes of the normalization of the
kinetic term, the deformation {\it does} change the central charge
of the Liouville field theory through the  Fradkin-Tseytlin counter
term \cite{Fradkin:1984pq}\cite{Fradkin:1985ys} $\delta Q \phi R$,
which vanishes on the flat Euclidean space we are using. The
non-compactness of the target-space, however, makes the deformation
exactly marginal by avoiding the $c$-theorem
\cite{Zamolodchikov:1986gt}.} This guarantees that there are no
singular terms that cannot be absorbed by the field re-definition in
the Liouville OPE from such deformation. More formally, one could
define  $\partial_\mu\phi \partial^\mu \phi$ as $4:L_{-1} V'_0
\bar{L}_{-1} V'_0:$ in the abstract Liouville field theory language,
and study the OPE. To evaluate the OPE among $\partial_\mu\phi
\partial^\mu \phi$, one can first investigate the OPE among the
Logarithmic primary operators:\begin{align} L_{-1} V'_0 (z) \cdot
\bar{L}_{-1} V'_0 (0) &\sim \frac{S(b)}{|z|^2} \cr L_{-1}V'_0(z)
\cdot L_{-1} V'_0 (0) &\sim \frac{1}{z^2} +
\frac{S(b)\log(\bar{z})}{z^2} \ ,
\end{align}
and so on. Note that $L_{-1} V'_0$ (or $\bar{L}_{-1} V'_0$) is
no-longer a left (right) moving primary operator but still is a
right (left) moving primary operator \cite{Zamolodchikov:2003yb}.
One can now see that the leading OPE singularity among
$\partial_\mu\phi \partial^\mu \phi$ is exactly the same as that for
the free scalar field theory, which means that the addition of the
term simply changes the normalization of the kinetic term of the
Liouville field, as in the free scalar field theory. The additional
logarithmic term should be renormalized by the Fradkin-Tseytlin
counter term, which is indeed necessary to keep the scale invariance
even in the Liouville theory with no deformation (e.g. Polyakov
regularization \cite{Polyakov:1981rd} gives $\lim_{w\to
z}\log|w-z|^2 = -2 \log |\rho(z)|^2$, where $\sqrt{g}R =
-4\partial\bar\partial \log |\rho|^2$).

Secondly, in the sigma model sector, we note  the fact that the
light-cone scalar is non-singular in its OPE, namely
$(X^1-X^0)(z)\cdot (X^1-X^0)(0) \sim 0 $ which implies
$e^{iu(X^1-X^0)}(z) \cdot e^{iv(X^1-X^0)}(0) \sim
e^{i(u+v)(X^1-X^0)}(0)$, suggests that there are actually no
additional singular contributions to the whole perturbation series.

Combining all these two sectors together, we have no hidden cut-off
dependence in the partition function (even with background charge),
and, therefore, we preserve the scale invariance under the
perturbation to all oder in $\lambda$.\footnote{The argument here
actually suggests that a broader class of non-conformal but scale
invariant field theories be obtained by choosing arbitrary
left-moving function $h(X^N) = h(X^1-X^0)$. \label{tt}} Of course,
some correlation functions are modified and operators acquire extra
anomalous dimension matrices, but they should be renormalized
independently of the beta function.

\section{Discussion}
In this paper, we have shown an example of scale invariant but
non-conformal quantum field theories in two dimension. From the
general argument \cite{Polchinski:1987dy}, such a theory should be
non-compact or non-unitary. In our case, the theory is both
non-compact and non-unitary. The former is due to the Liouville
direction and the latter is due to the time-like direction in the
sigma model. Indeed, the correlation function of the trace of the
(improved) energy-momentum tensor
\begin{equation}
\left\langle \Theta^{\mu \ \dagger}_{\ \mu} \Theta^{\mu}_{\ \mu}
\right \rangle
\end{equation}
vanishes due to the charge conservation,\footnote{The argument is as
follows. We set $\Theta^{\mu}_{\ \mu} =  -
\frac{4}{b}\partial\bar{\partial}\phi + 4\pi \mu e^{2b\phi}$. The
perturbative computation with respect to $\lambda$ should be done at
$\lambda = 0$ because of the charge conservation. Then, the
correlation function vanishes due to the Liouville equation of
motion.} while the trace itself does not vanish as we have seen in
the previous section. The failure of the proof in
\cite{Polchinski:1987dy} here is due to this non-unitary nature of
the correlation functions, which manifests itself as the lack of
reflection positivity.

Although our model might have no physical significance as a two
dimensional field theory because of the lack of the unitarity, it
may have some applications in string theory, where the world-sheet
theory needs not be unitary as long as ghosts are removed by the
BRST constraint. From the viewpoint of the string worldsheet
perturbation theory, this kind of exactly marginal but non-conformal
deformation would be quite dangerous because it induces a
world-sheet conformal anomaly, and it would lead to a potential
swampland from the target-space viewpoint. Fortunately, the central
charge of the Liouville sector is $c_{\phi}=1+6(b+b^{-1})^2 \ge 25$,
and the two extra dimensions for $X^1$ and $X^0$ make it difficult
to embed our models in the critical string theory.

This dangerous situation could occur in the super-critical string
theory (see e.g. \cite{Hellerman:2006nx}\cite{Hellerman:2007fc} and
references therein), where we can introduce the time-like linear
dilaton as well to reduce the central charge of the $X^0$ scalar as
$c_{X^0}=1-6(\beta - \beta^{-1})^2$ where $\beta$ is the slope of
the time-like linear dilaton. The world-sheet perturbation
$e^{(2i\omega-(\beta-\beta^{-1})) X^0 - 2ikX^1} \partial \phi
\bar{\partial}\phi$ could be an exactly marginal deformation
(Liouville wave) under the condition $
-\frac{(\beta-\beta^{-1})^2}{4}-\omega^2 + k^2 = 0$. If the
perturbation is exactly marginal, such a background would be
inconsistent as a string background although the scale invariance is
intact. It would be very interesting to see whether the Liouville
wave deformation is possible within the super-critical string theory
and investigate a possibly critical consequence of such a dangerous
deformation.

\section*{Acknowledgements}
One of the authors Y.~N. would like to dedicate the paper to the
memory of Alyosha Zamolodchikov, who shared the author with his
enthusiasm in the Liouville field theory and his higher equations of
motion. The research of Y.~N. is supported in part by NSF grant
PHY-0555662 and the UC Berkeley Center for Theoretical Physics.
C.~M.~Ho acknowledges the support from the Croucher Foundation and
Berkeley CTP.

\end{document}